\documentclass[twocolumn,aps]{revtex4}
\usepackage{graphicx}
\usepackage{dcolumn}
\usepackage{bm}
\usepackage[bf]{subfigure}
\usepackage{rotating}
\usepackage{amssymb}

\begin{document}
%
\title{Trees = Networks ?!?}

\author{L.\ da\ F.\ Costa}
\author{F.\ A.\ Rodrigues}
\affiliation{Instituto de F\'{\i}sica de S\~{a}o Carlos,
Universidade de S\~{a}o Paulo, PO Box 369, 13560-970, phone +55 16
3373 9858, FAX +55 16 3371 3616, S\~{a}o Carlos, SP, Brazil }

\begin{abstract}
This work addresses the intrinsic relationship between trees and
networks (i.e. graphs).  A complete (invertible) mapping is presented
which allows trees to be mapped into weighted graphs and then
backmapped into the original tree without loss of information.  The
extension of this methodology to more general networks, including
unweighted structures, is also discussed and illustrated.  It is shown
that the identified duality between trees and graphs underlies several
key concepts and issues of current interest in complex networks,
including comprehensive characterization of trees and community
detection. For instance, additional information about tree structures
(e.g. phylogenetic trees) can be immediately obtained by taking into
account several off-the-shelf network measurements --- such as the
clustering coefficient, degree correlations and betweenness
centrality.  At the same time, the hierarchical structure of networks,
including the respective communities, becomes clear when the network
is represented in terms of the respective tree. Indeed, the
network-tree mapping described in this work provides a simple and yet
effective means of community detection.
\end{abstract}


\maketitle

\vspace{0.5cm}
\emph{`All things must change to something new, to
something strange.'
(H. W. Longfellow)}

\section{Introduction}

Trees and general networks (or graphs) are seemingly rather distinct
structures, the former characterized by a strict hierarchical
organization and absence of cycles, while the latter are generally
tangled and involve several cycles.  Trees are frequently associated
to systems underlain by hierarchies, branching and taxonomies ---
including but not being limited to phylogenetics, vascularization,
hydrography, and even neuronal shapes.  Because of their deceptive
simplicity (especially absence of cycles), trees are rarely
investigated by considering topological measurements other than the
number of levels, branches and symmetry (e.g~\cite{villasboas2007btc,
villasboas2008cmt}).  On the other hand, networks or graphs are
seemingly more general discrete structures, underlying a large number
of natural complex systems ranging from the Internet to the brain
(e.g.~\cite{Costa08:surveyapp, Boccaletti06:PR}).  A wealthy of
measurements has been proposed and applied to characterize the
topology of complex networks~\cite{Costa:survey}, which can be found
in two types: weighted and unweighted, with the latter being a
specific instance of the former~\cite{Costa:survey}.

\begin{figure*}
  \centerline{\includegraphics[width=11cm]{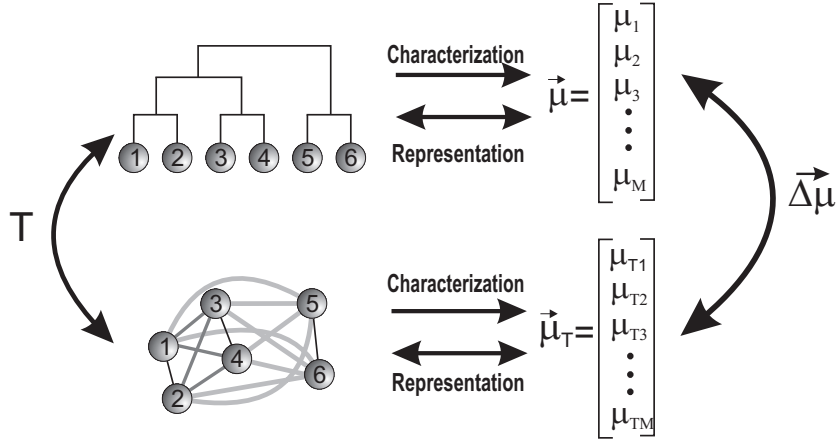}}
  \caption{The transformation of the network into a tree can provide
              additional information for its characterization.}
  \label{Fig:transf}
\end{figure*}

Though trees and weighted networks have been treated largely in
independent fashion, there is an intrinsic and important relationship
(duality) between these two seemingly different types of discrete
structures that can lead to a number of interesting implications of
theoretical and practical significance. To a limited extent, the
duality tree-network has been sporadic and indirectly explored in some
works, especially those aimed at mapping the hierarchical/modular
structure of complex networks (e.g.~\cite{Newman04:EPJ,
Newman04:PRE}). The understanding of trees as networks, however, has
received even less attention from the complex network community. Yet,
provided a tree can be meaningfully mapped into a network, a series of
comprehensive measurements can be immediately obtained which can
potentially lead to a better understanding and modeling of the
original trees, as illustrated in
Figure~\ref{Fig:transf}~\cite{Costa:survey}.  Given a tree and its
respective limited set of measurements $\vec{\mu}$, it is mapped into
a respective weighted graph which can then be characterized by a
richer set of features $\vec{\mu_T}$.  The joint consideration of
these two sets of measurements, as well as their difference
$\vec{\Delta \mu}$, provide a very comprehensive resource for
analysing and classifying the original tree.
Figure~\ref{Fig:transf}~\cite{Costa:survey} also illustrates that the
mapping from a discrete structure into a respective set of
measurements can be invertible (representation) or not invertible
(characterization).

One of the main objectives of the present work is to describe a
perfect duality between trees and respective weighted networks.  It is
also shown that such results can be extended to more general networks,
i.e. those which can not be obtained by transformations of trees as
well as unweighted structures. The implications of such a duality are
many from both theoretic and applied points-of-view.

This work starts by describing how trees can be mapped into respective
weighted networks and then backmapped without any loss of information.
A simple algorithm is presented and illustrated for that finality.  We
subsequently address the situation involving more general unweighted
networks.

\section{Basic Concepts and Methods}

\begin{figure}[t]
  \centerline{\includegraphics[width=0.95\columnwidth]{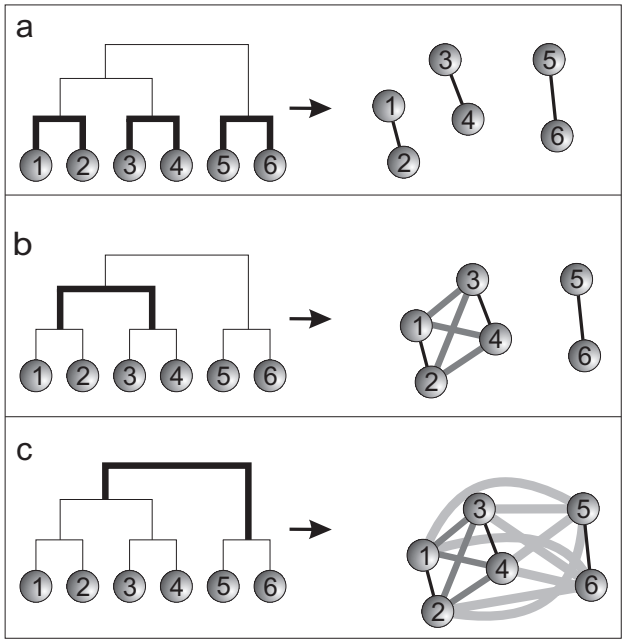}}
  \caption{Example of the adopted methodology to transform a tree into a
            network. }
  \label{Fig:ex_tree-net}
\end{figure}

The transformation of a tree into a network, henceforth called
\emph{tree-net mapping}, is relatively simple and can be performed as
follows.  First, as illustrated in Figure~\ref{Fig:ex_tree-net}, the
nodes (leaves) of the same branch (from the bottom of the tree
upwards) are connected with weights equal to one. Then, the nodes of
the groups at the next level of the tree are connected with the
vertices of other groups at the same level through edges of weight
2. Every vertex in each branch is connected to every vertex of the
other groups (a clique). This process is repeated until we reach the
top of the tree. The last connections will have the maximum weight
$2H$, equal to twice the height of the three.  Therefore, the
resulting weighted networks will have their nodes interconnected in
order to reflect the branching pattern of the original tree.  The
weighting scheme adopted above is justified in order to reflect the
distances between nodes along the tree connectivity.  Weighted
networks that can be derived from trees by the above methodology are
henceforth called \emph{tree-ancestered networks} and abbreviated as
TAN.

\begin{figure}[t]
  \centerline{\includegraphics[width=0.9\columnwidth]{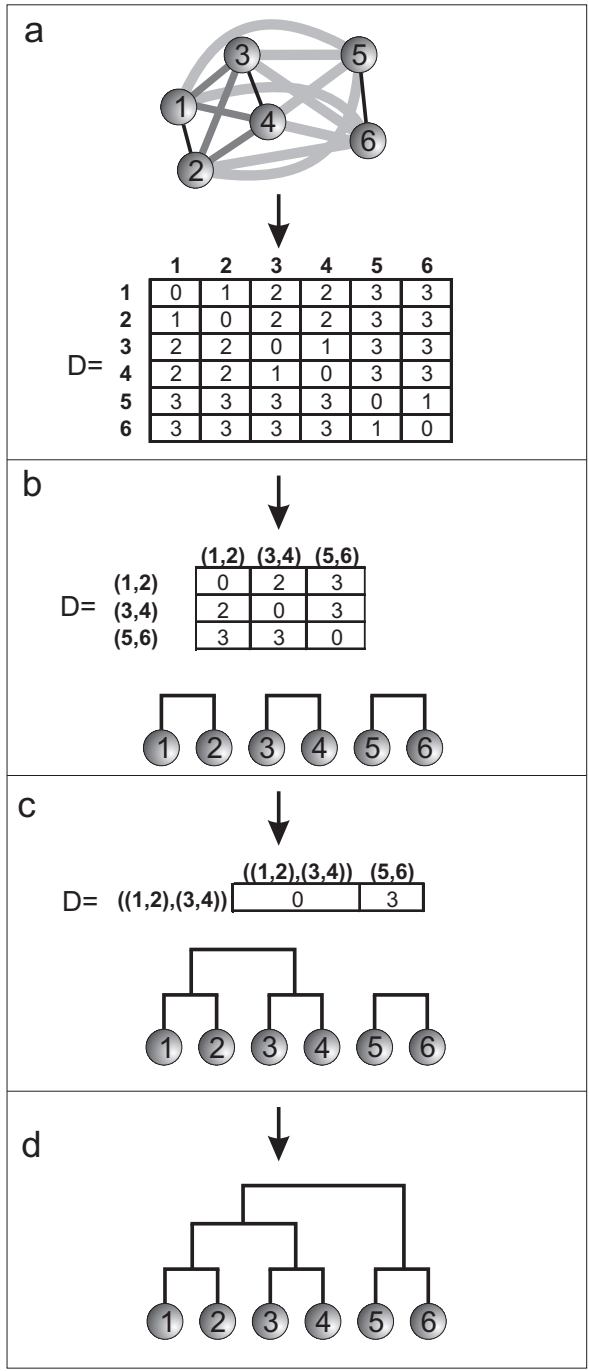}}
  \caption{Example of the methodology to transform a network into a tree.
             Note that the columns and rows of the distance matrix are
             joined at each step, while the new elements are equal to the
             average of the grouped columns.}
  \label{Fig:ex_net-tree}
\end{figure}

The second type of transformation considered in this work, the
so-called \emph{network-tree mapping}, considers an
agglomerative-thresholding method in order to backmap a TAN into the
original tree. This can be achieved by using the following simple
algorithm (please refer to Figure~\ref{Fig:ex_net-tree}):

\begin{enumerate}
  \item The weakest edges $(i,j)$ are determined;
  \item Each respective group of vertices which belong to a connected
         component considering that edges weight are subsumed in
         the weighted matrix, with the new columns corresponding to
         the average between the respective joined columns.
\end{enumerate}

These steps are repeated until the original tree is obtained.
Figure~\ref{Fig:ex_net-tree} illustrates an application example of
such methodology.  This backmapping can be shown to be lossless, in
the sense that given a TAN, its original respective tree can be
perfectly recovered.

While the previous pair of algorithms (i.e. tree-net and net-tree),
allowing the transformation of any tree into a respective TAN and then
backmapping it into the original tree, define a perfect duality
between those two types of discrete structures, it is also interesting
to consider the more general case of transforming any network,
including unweighted graphs, into trees.  This transformation is
intrinsically related to community detection in complex networks
(e.g.~\cite{Hopcroft04,zhou2003ddi}).  In the current work, instead of
trying to obtain a respective tree corresponding to a unweighted
network, we first transform it into a weighted network by considering
the shortest distances between each pair of nodes~\cite{costa2008scs,
bagrow2008pcn, andrade2006npc}.  The resulting network is fully connected and the
edge weights between every pair of nodes $i$ and $j$ is equal to their
respective shortest path distance.  We considered the following
algorithm to transform a generic unweighted network into the
respective tree:
\begin{enumerate}
  \item Compute the distance matrix of the network~\cite{floyd1962asp}.
  \item Compute the Euclidean distance between every pair of vertices
           considering their respective columns in the distance matrix.
  \item Identify the edges with the shortest Euclidean distance
           (i.e. the smallest entries in the current weight matrix).
  \item Join the edges which belong to respective connected components.
  \item Merge the respective vertices and put in the new column and row
                respective weight averages.
\end{enumerate}

\begin{figure}[t]
  \centerline{\includegraphics[width=0.9\columnwidth]{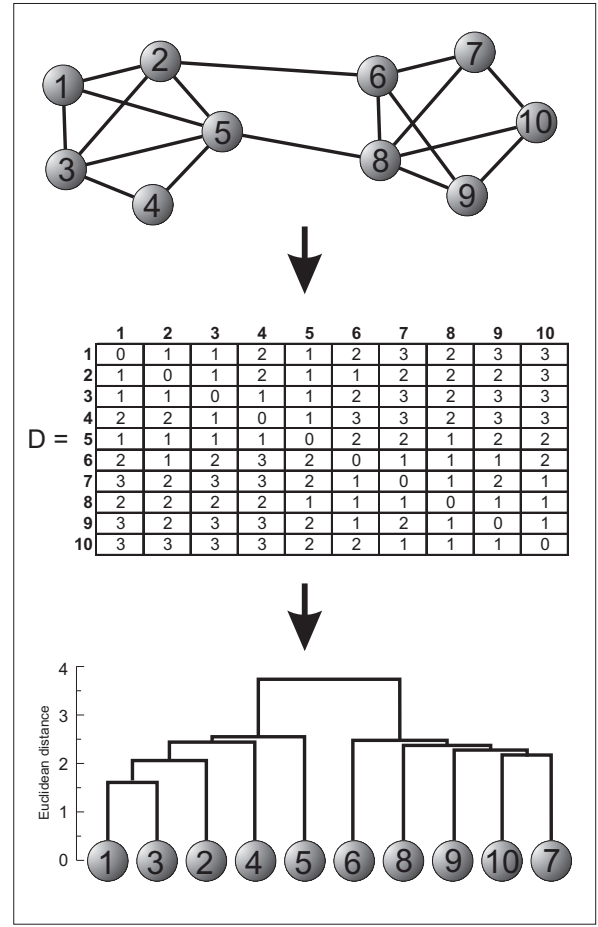}}
  \caption{A generic, unweighted network can be transformed into a
             weighted network with respect to the distances between
             all nodes and then into a tree by using the hierarchical
             agglomerative method proposed in the current work.}
  \label{Fig:net_tree}
\end{figure}

Figure~\ref{Fig:net_tree} illustrates the described methodology for a
simple unweighted network.  It is clear from this example that the two
communities in the original network were clearly distinguished as the
two main branches in the respectively obtained tree.  It is important
to observe that the transformation of this tree into a network by
using the first algorithm presented in this work will yield a network
with fully connected communities (i.e. cliques) considering
subsequence weights (i.e. distances).  Though the generic
network-tree-network sequence of transformations is typically not
invertible, the therefore obtained weighted network represents an
interesting structure which emphasizes the communities hidden in the
original unweighted network.  By comparing the original, unweighted
network, and the network obtained by the network-tree-network
transformation, it becomes possible to identify the missing links
among the nodes in each detected community.

\section{Conclusion}

Trees and networks (or graphs) correspond to two of the most important
and frequently discrete structures in physics, biology and computer
science.  Though traditionally treated as quite different structures,
they are actually closely inter-related as shown in the present
work. First, trees can be transformed into weighted networks (TANs)
which can then be backmapped into the original tree by using simple
respective algorithms.  Such a perfect duality is important at least
because of the two reasons: (i) by transforming trees into networks,
it becomes possible to characterize their topological features in
terms of a wealthy of off-the-shelf complex networks
measurements~\cite{Costa:survey}; and (ii) the transformation of
networks into trees is intrinsically related to community finding and
the hierarchical organization of networks.

We have also shown that even generic, unweighted networks can be
mapped into trees.  This can be achieved by first transforming the
original unweighted network into a weighted network, whose weights
correspond to the distances between the nodes in the original
network. The respective tree is then obtained by progressive merging
of nodes according to a simple agglomerative algorithm.  The
comparison between this network and the original, unweighted network
has potential for highlighting the missing links amongst the nodes
in each community.

While the current work has been limited to motivating and presenting
the three algorithms for transforming between trees and networks, as
well as the perfect duality between trees and respectively transformed
weighted networks, future developments can be performed in order to
apply these results to real-word and theoretical complex networks.  It
would also be interesting to compare the performance of the generic
unweighted network to tree transformation as a potentially effective
and computationally parsimonious method.

\begin{acknowledgments}
Luciano da F. Costa thanks CNPq (308231/03-1) and FAPESP (05/00587-5)
for sponsorship.  Part of this work was developed during the author's
Visiting Scholarship at St. Catharine's College, University of
Cambridge, UK.
\end{acknowledgments}

\bibliographystyle{apsrev}
\bibliography{treenet}

\end{document}